\documentclass[twocolumn,letter]{jpsj3} 
\usepackage{txfonts}

\title{
Spin-Hall Effect and Diamagnetism of Dirac Electrons}

\author{Yuki Fuseya\thanks{E-mail address: fuseya@mp.es.osaka-u.ac.jp}
\thanks{Present adress: Department of Applied Physics and Chemistry, The University of Electro-Communications, Chofu, Tokyo 182-8585} 
 $^{1} $,
 Masao Ogata $^{2}$,
 and 
 Hidetoshi Fukuyama$^{3}$
}
\inst{
$^{1}$ Department of Materials Engineering Science, Osaka University, Toyonaka, Osaka 560-8531\\
$^{2}$ Department of Physics, University of Tokyo, 7-3-1 Hongo, Bunkyo-ku, Tokyo 113-0033\\
$^{3}$ Department of Applied Physics and Research Institute for Science and Technology, Tokyo University of Science, Kagurazaka, Shinjuku-ku, Tokyo 162-860
}

\abst{
Spin-Hall conductivity (SHC) of fully relativistic ($4\times 4$ matrix) Dirac electrons is studied based on the Kubo formula aiming at possible application to bismuth and bismuth-antimony alloys. 
It is found that there are two distinct contributions to SHC, one only from the states near the Fermi energy and the other from all the occupied states.
The latter remains even in the insulating state, i.e., when the chemical potential lies in the band-gap, and turns to have the same dependences on the chemical potential as the orbital susceptibility (diamagnetism), a surprising fact.
These results are applied to bismuth-antimony alloys and the doping dependence of the SHC is proposed.
}

\kword{spin-Hall effect, Dirac electron, spin-orbit interaction, Kubo formula, diamagnetism, bismuth}

\usepackage{amsmath,bm,mathrsfs,color}
\renewcommand{\Im}{{\rm i}}

\newcommand{\D}{\Delta}

\newcommand{\bk}{\bm{k}}
\newcommand{\bp}{\bm{p}}

\newcommand{\scr}[1]{\mathscr{#1}}
\newcommand{\ve}{\varepsilon}
\newcommand{\tve}{\tilde{\varepsilon}}
\newcommand{\sgn}{{\rm sgn}}

\begin{document}
\maketitle
	%
	The spin-orbit (SO) interaction has been known to lead to non-trivial and intriguing phenomena such as anisotropic magnetoresistance\cite{Thomson1857,Binasch1989} and anomalous Hall effect\cite{Karplus1954,Nagaosa2010}.
	More recent example is the spin-Hall effect (SHE) which indicates the generation of spin-current perpendicular to an external electric field\cite{Nagaosa2008,Dyakonov2008,Murakami2003,Sinova2004,Murakami2004,Kontani2007,Kontani2008}.
	One of the interesting features of SHE is the possibility of finite contributions even in the insulating states called as spin-Hall insulators.\cite{Murakami2004}
	Basically, materials with strong SO interaction should be the stage to observe such interesting features associated with SHE.
	In this context, bismuth and its alloys with antimony will be one of the best candidates since SO interactions are known to play crucial roles in these systems\cite{Fukuyama1970}.

	Crystalline bismuth is basically cubic but with slight distortions (Peierls distortions) along the diagonal directions leading to semi-metallic electronic states with small number of electrons ($L$-points) and holes ($T$-point)\cite{Dresselhaus1971}.
	Its electronic state has very characteristic features of anisotropy of small effective masses and large $g$-factors ($g \sim $ 100-1000)\cite{Smith1964,Zhu2011}, resulting from both small band-gap and strong SO interaction.
	Such characteristic electronic states are described by a $4\times 4$ matrix Hamiltonian (conduction and valence bands with spin degrees of freedom) around $L$ or $T$ points in the $\bk \cdot \bp$ representation.
	This Hamiltonian, which may be called Wolff Hamiltonian, describes the low-energy properties of bismuth quite well, and turns out to be essentially the same as that for Dirac Hamiltonian except with the anisotropy of velocity here,\cite{Wolff1964} because the space in solids is in general anisotropic in contrast to true vacuum.
	(The Wolff Hamiltonian reduces to well-known Dirac Hamiltonian if spatial anisotropy of velocity is ignored, but with different velocity.)

	Bismuth is known for its very large diamagnetism with the particular feature of having maximum in the insulating states (by changing the chemical potential with alloying) whose origin is now understood based on this Wolff Hamiltonian as due to inter-band effects of magnetic field in the presence of strong SO.\cite{Fukuyama1970}
	This inter-band effect also affects the off-diagonal transport coefficient, but not the diagonal transport, in general\cite{Kubo1970}.
	The inter-band effect in the off-diagonal transport coefficient and that in the diamagnetism should be related to each other in some way, but not clearly understood yet --- a longstanding problem\cite{Kubo1970}. 
	%
	%
	
	%
	In this paper, we study the SHE based on the Kubo formula for the isotropic Wolff Hamiltonian\cite{Fuseya2009,Fuseya2012}, i.e., the fully relativistic Dirac Hamiltonian, not only in insulating but also in conducting states on an equal footing.
	We find that there are two distinct contributions to SHE, one from the states near the Fermi energy and the other from all the occupied state.
	The latter turns to have the same dependences on the chemical potential as the orbital diamagnetism indicating that there is a close relationship between the SHE and diamagnetism in bismuth. 
	%
	
	%

	We start from a simple one electron Hamiltonian with the SO interaction
	\begin{align}
		\scr{H}=\frac{p^2}{2m}+V + \frac{\nabla^2 V}{8m^2 c^2}+ \frac{1}{4m^2 c^2}
		\bp \cdot \left( \bm{\sigma}\times \bm{\nabla} V \right),
		\label{CB}
\end{align}
	as an effective model of electrons at $L$ point in bismuth\cite{Cohen1960}.
	Here the Pauli matrix $\bm{\sigma}$ corresponds to the real spin of electrons, $V$ is the crystal potential, and the last term expresses the SO interaction.
	This effective Hamiltonian can be transformed into in an essentially identical form to the Dirac Hamiltonian as is shown by Wolff\cite{Wolff1964}.
	Here we discuss the isotropic case of the Wolff model\cite{Fuseya2009,Fuseya2012}:
\begin{align}
	\scr{H}=
	\begin{pmatrix}
		\D & \Im \gamma \bk \cdot \bm{\sigma} \\
		-\Im \gamma \bk \cdot \bm{\sigma} & -\D
	\end{pmatrix},
	\label{Hamiltonian}
\end{align}
	where $2\D$ is the band gap.
	The SO interaction is included in the matrix elements $\gamma$, and the parabolic part in eq. (\ref{CB}) is discarded since it is negligibly small compared to the SO term in case of bismuth\cite{Smith1964,Zhu2011}.
	Originally, the matrix elements are anisotropic,
	%
	but we have assumed that all matrix elements are equal in order to make our arguments as simple and transparent as possible  in eq. (\ref{Hamiltonian}).
	%
	%
	The eigen energy of this Hamiltonian is $\pm E_k= \pm\sqrt{\gamma^2 k^2 + \D^2}$.
	The velocity operator is defined by $\bm{v} = \partial \scr{H} /\partial \bk$.

	As in the Dirac theory, the magnetic moment of electrons, $\bm{\mu}_{\rm e}$, can be determined as the coefficient of $\bm{\sigma} \cdot \bm{B}$ term.
	Then we have\cite{Fuseya2012}
	\begin{align}
		\bm{\mu}_{\rm e}=\frac{g^{*}\mu_{\rm B}}{2}
		\begin{pmatrix}
			-\bm{\sigma} & 0\\
			0 & \bm{\sigma}
		\end{pmatrix},
		\label{moment}
\end{align}
	where $g^{*}=2m_{\rm e}\gamma^{2}/\D$ is the effective g-factor, $m_{\rm e}$ is the free electron mass, and $\mu_{\rm B}=e/2m_{\rm e}c$ the Bohr magneton.
	Note that $g^{*}$ is given as the reciprocal of the effective cyclotron mass, $g^{*}=2m_{\rm e}/m_{\rm c}^{*}$, where $m_{\rm c}^{*}=\D / \gamma^{2}$.
	It should be emphasized here that the sign of the magnetic moment are opposite between the conduction and valence band for the Dirac electrons.
	Correspondingly, the spin-velocity operator, $v_{{\rm s}i}$ ($i=x, y$), which can be defined by the velocity of the magnetic moment along the $z$-direction, is given as\cite{Fuseya2012}
	\begin{align}
		v_{{\rm s}i} 
		=
		\frac{\mu_{{\rm e}z}v_{i}}{\mu_{\rm B}}
		= - \frac{\Im m_{\rm e}\gamma^{3}}{\D}
		\begin{pmatrix}
		0 & \sigma_z \sigma_i \\
		\sigma_z  \sigma_i & 0
		\end{pmatrix}.
\end{align}
	This is an Hermitian operator.

	The spin-Hall conductivity (SHC) is given as a linear response of the spin-velocity operator to the electric field on the basis of the Kubo formula:
	\begin{align}
		\sigma_{{\rm s}yx} &= \frac{1}{\Im \omega}
		\left[
			\Phi_{{\rm s}yx}(\omega + \Im \delta ) - \Phi_{{\rm s}yx}(0 + \Im \delta )
		\right],
		\\
		\Phi_{{\rm s}yx}(\Im \omega_\lambda)&=
		-e T \sum_{n, \bk} {\rm Tr}\left[
		 \scr{G}(\Im \tilde{\ve}_n )v_{{\rm s}y} \scr{G}(\Im \tilde{\ve}_{n-}) v_x
		\right],
		\label{Phi6}
\end{align}
	where $\ve_{n-} = \ve_n - \omega_\lambda$, and $\scr{G}(\Im \tilde{\ve}_n) = \left( \Im \tilde{\ve}_n - \scr{H} \right)^{-1}$ is the Green function of eq. (\ref{Hamiltonian}) with $\Im \tilde{\ve}_n = \Im \ve_n + \Im \Gamma \sgn (\ve_n)$.
	Here we have introduced phenomenologically a quasiparticle damping rate $\Gamma$ as an imaginary part of the self energy.

	After some straightforward calculations, we have
	\begin{align}
		\Phi_{{\rm s}yx}(\Im \omega_\lambda) = -e T 
		\sum_{n, \bk}
		\frac{4m_{\rm e}\gamma^4 \left( \tve_n - \tve_{n-}\right)}
	{\left( \tve_n^2 + E_k^2 \right) \left( \tve_{n-}^2 + E_k^2 \right)}.
	\label{Phi}
\end{align}
	We perform $n$-summation by the standard analytic continuation technique, and carry out the momentum integration.
	Then we obtain the following formula of the SHC
	\begin{align}
		\sigma_{{\rm s}yx} &= -\frac{ em_{\rm e} |\gamma|}{4\pi^{2}} \left( K_{{\rm s}yx}^{\rm I} + K_{{\rm s}yx}^{\rm II} \right),
		\label{Csxy}
		 \\
		K_{{\rm s}yx}^{\rm I} &=	
		\int_{-\infty}^\infty \!\!d\ve
 		f'(\ve)
		\left[
		\frac{\sqrt{\ve_{+}^2 -\D^2}}{\ve} 
		-\frac{\sqrt{\ve_{-}^{2} -\D^2}}{\ve}
		\right],
		\label{K1}
		\\
		K_{{\rm s}yx}^{\rm II} &=
		\int_{-\infty}^\infty \!\! d\ve f(\ve)
		\left[
		\frac{1}{\sqrt{\ve_{+}^2 -\D^2}}
		-\frac{1}{\sqrt{\ve_{-}^2 -\D^2}}
		\right],
		\label{K2}
\end{align}
	where $\ve_{\pm}=\ve \pm \Im \Gamma$, $f(\ve)$ and $f'(\ve)$ are the Fermi distribution function, whose energy is measured from the chemical potential $\mu$, and its derivative, respectively.
	(The branch cut of the square root is taken along the positive real axis.)
	Note that the diagonal spin conductivity, $\sigma_{{\rm s}xx}$ is exactly zero in the case of the present Hamiltonian.

	In deriving (\ref{Csxy})-(\ref{K2}), we have four contributions coming from the contours $C_{1\sim4}$ in the complex $z$ plane:
	$C_{1}$ ($C_{2}$) is the contour from $-\infty$ to $+\infty$ (from $+\infty$ to $-\infty$) along just above (below) the horizontal line ${\rm Im}\, z = \omega_{\lambda}$, and 
	$C_{3}$ ($C_{4}$) is the contour from $-\infty$ to $+\infty$ (from $+\infty$ to $-\infty$) along just above (below) the horizontal line ${\rm Im}\, z = 0$.
	$K_{{\rm s}yx}^{\rm I}$ in eq. (\ref{K1}) originates from the contribution of $C_{2}+C_{3}$.
	It has a functional form 
	\begin{align}
		-\frac{1}{2\pi \Im} \int_{-\infty}^{\infty}\!\!d\ve f(\ve)\left[
		G^{\rm R}(\ve)G^{\rm A}(\ve-\omega)-G^{\rm R}(\ve+\omega)G^{\rm A}(\ve)
		\right], 
		\label{C23}
	\end{align}
	where $G^{\rm R (A)}$ is the retarded (advanced) Green function.
	Confining ourselves to the static response, we need only $\omega$-linear term.
	Shift of the variable $\ve \to \ve - \omega$ in the second term of eq. (\ref{C23}) leads to
	%
$
		\frac{\Im \omega}{2\pi} \int_{-\infty}^{\infty}\!\! d\ve
		\left( \frac{d f(\ve)}{d\ve} \right)
		G^{\rm R}(\ve) G^{\rm A}(\ve)
$
	%
	in the limit of $\omega \to 0$.
	Therefore, the contribution of $K_{{\rm s}yx}^{\rm I}$ is only from the states near the Fermi energy.
	This is similar to the contribution in the transport properties, which may be called ``transport contribution''.
	%
	%
	
	The second term, $K_{{\rm s}yx}^{\rm II}$, in eq. (\ref{K2}) originates from the contribution of $C_{1}+C_{4}$.
	It has a functional form of $\omega f(\ve) \left[G^{\rm R}(\ve) G^{\rm R}(\ve) - G^{\rm A}(\ve) G^{\rm A}(\ve)\right]$.
	This is similar to the contribution in the thermodynamical quantities such as thermodynamic potential, which may be called ``thermodynamic contribution''.
	In this case, the factor $f(\ve)$ sums up the contributions from all the states below the Fermi energy.

	For the clean limit, $\Gamma \to 0$, at zero temperature, eqs. (\ref{K1}) and (\ref{K2}) can be expressed in the simple forms:
	\begin{align}
		-K_{{\rm s}yx}^{\rm I} &=
		\left\{
		\begin{array}{ll}
		\displaystyle \frac{2\sqrt{\mu^2 -\D^2}}{|\mu|} & (|\mu|>\D) \\
		0 & (|\mu|<\D)
		\end{array}
		\right.,
		\\
		-K_{{\rm s}yx}^{\rm II} &=
		\left\{
		\begin{array}{ll}
			\displaystyle 2\ln \left(\frac{2E_{\rm c}}{|\mu| + \sqrt{\mu^2-\D^2}}\right) & (|\mu|>\D) \\
			\displaystyle 2 \ln \left( \frac{2E_{\rm c}}{\D}\right) & (|\mu|<\D)
		\end{array}
		\right.,
\end{align}
	where $E_{\rm c}$ is the energy cutoff for the integration, and we discarded $\mathcal{O}(\D^2 /E_{\rm c}^2)$-term.
	%
	
	%
\begin{figure}
\begin{center}
\includegraphics[width=8cm]{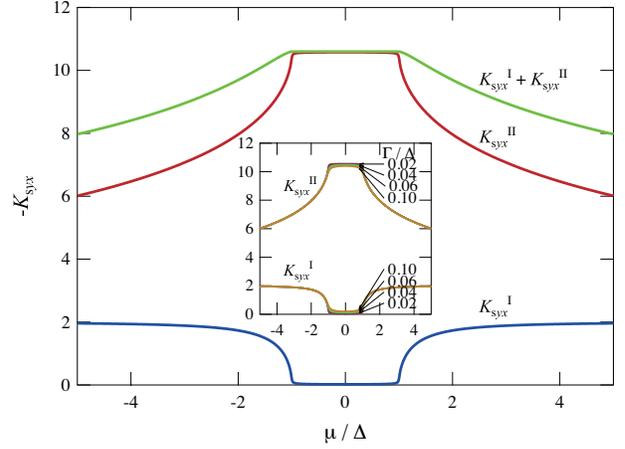}
\end{center}
\caption{(Color online) Chemical potential dependence of the SHC ($K_{{\rm s}yx}^{\rm I, II}$) for $\Gamma/\D=0.01$ and $E_{\rm c}/\D = 100$.
The inset shows the plot of $K_{{\rm s}yx}^{\rm I, II}$ for different damping rates: $\Gamma /\D =0.02, 0.04, 0.06, 0.10$.
}
\label{f1}
\end{figure}
	
	The chemical potential dependences of $K_{{\rm s}yx}^{\rm I}$ and $K_{{\rm s}yx}^{\rm II}$ are shown in Fig. \ref{f1} for which the values of $\Gamma /\D$ and $E_{\rm c}/\D$ are chosen to 0.01 and 100, respectively.
	%
	%
	$K_{{\rm s}yx}^{\rm II}$ gives the dominant contribution for the SHE.
	Furthermore, $K_{{\rm s}yx}^{\rm II}$, and so the total $\sigma_{{\rm s}yx}$ become maximum in the insulating region, $|\mu|<\D$, and decays logarithmically for $|\mu|>\D$ although the carrier number increases.
	$\Gamma$-dependences of $K_{{\rm s}yx}^{\rm I, II}$ are shown in the inset of Fig. \ref{f1}.
	The small $\Gamma$-dependence of both $K_{{\rm s}yx}^{\rm I}$ and $K_{{\rm s}yx}^{\rm II}$ (and so the total $\sigma_{{\rm s}yx}$) means that the present SHE corresponds to the so-called intrinsic SHE\cite{Murakami2003,Sinova2004,Murakami2004,Kontani2007,Kontani2008}.


	Now, let us discuss the inter-band effect on the SHE.
	We can divide $\Phi_{{\rm s}yx}(\Im \omega_{\lambda})$ into the contributions from the intra- or inter-band effect using the formula:\cite{Fuseya2012} 
	$\Phi_{{\rm s}yx} (\Im \omega_{\lambda})=\sum_{\alpha \beta}\Phi_{{\rm s}yx}^{\alpha \beta} (\Im \omega_{\lambda})$ with
	\begin{align}
		\Phi_{{\rm s}yx}^{\alpha \beta} (\Im \omega_{\lambda})=
		-eT \sum_{n, \bk}\left[
		\langle \psi_{\alpha} | v_{{\rm s}y} | \psi_{\beta} \rangle
		\langle \psi_{\beta} | v_{x} | \psi_{\alpha} \rangle
		\scr{G}_{\alpha}(\Im \tilde{\ve}_{n})\scr{G}_{\beta}(\Im \tilde{\ve}_{n-}) 
		\right],
		\label{intra-inter}
\end{align}
	where $\alpha, \beta =\pm$ denote the conduction ($+$) or valence ($-$) band, $\psi_{\pm}$ are their wave functions, and $\scr{G}_{\pm}(\Im \ve_{n})=\left[ \Im \ve_{n} \mp E_k \right]^{-1}$.
	This formula is equivalent to eq. (\ref{Phi6}).
	The inter-band contribution is from $\Phi_{{\rm s}yx}^{+-}$ and $\Phi_{{\rm s}yx}^{-+}$, where the current and spin current vertices connect the conduction and valence bands.
	By explicit calculation of eq. (\ref{intra-inter}), we find that the SHC originates only from the inter-band part, namely,
	\begin{align}
		\sigma_{{\rm s}yx}^{\rm intra} &= \sigma_{{\rm s}yx}^{++}+\sigma_{{\rm s}yx}^{--} =0
		\\
		\sigma_{{\rm s}yx}^{\rm inter}&= \sigma_{{\rm s}yx}^{+-}+\sigma_{{\rm s}yx}^{-+}
		=
		-\frac{em_{\rm e}|\gamma |}{4\pi^{2}} 
		\left( K_{{\rm s}yx}^{\rm I} + K_{{\rm s}yx}^{\rm II} \right),
\end{align}
	where 
	$
		\sigma_{{\rm s}yx}^{\alpha \beta}=
		\left[\Phi_{{\rm s}yx}^{\alpha \beta}(\omega + \Im \delta)
		-\Phi_{{\rm s}yx}^{\alpha \beta}(0+ \Im \delta) \right]/\left( \Im \omega \right).
	$
	%
	%
	%
	
	The unusual properties of $\sigma_{{\rm s}yx}$ can be understood by comparing with $\sigma_{xx}$ and $\sigma_{yx}$\cite{Fuseya2009} in terms of intra- and inter-band contributions.
	In the case of the diagonal conductivity, $\sigma_{xx} (\omega \to 0)$, the dominant contribution comes from the intra-band part, and it is suppressed as $\Gamma$ increases ($\sigma_{xx}^{\rm intra}\propto \Gamma^{-1}$). 
	In the case of the Hall conductivity, $\sigma_{yx} (\omega \to 0)$, the dominant contribution also comes from the intra-band part, although there is a finite contribution from the inter-band part, which remains finite for $\Gamma \to 0$ and exhibits anomalous properties similar to that of the orbital susceptibility.\cite{Fuseya2009}
	The intra-band part is suppressed as $\sigma_{yx}^{\rm intra}\propto \Gamma^{-2}$, whereas the inter-band part is not. 
	In the case of the SHC, which is completely inter-band effect, the $\Gamma$-dependence of $\sigma_{{\rm s}yx}(\omega \to 0)$ is quite small as shown in Fig. \ref{f1} inset.
	From these results, we think that the $\Gamma$-dependence of the inter-band contribution will be generally small.
	%

	%
	Next, we discuss the relationship between $\sigma_{{\rm s}yx}$ and large diamagnetic susceptibility, $\chi$.
	Surprisingly, we find that the obtained $K_{{\rm s}yx}^{\rm II}$ has exactly the same $\mu$-dependence as $\chi$. 
	In the limit of weak magnetic field, $\chi$ is calculated by the simple but exact formula\cite{Fukuyama1971}
	\begin{align}
		\chi &= \frac{e^2}{c^2}T\sum_{n, \bk}
		{\rm Tr}\left[
		\scr{G}(\Im \tilde{\ve}) v_{x}
		\scr{G}(\Im \tilde{\ve}) v_{y}
		\scr{G}(\Im \tilde{\ve}) v_{x}
		\scr{G}(\Im \tilde{\ve}) v_{y}
		\right].
	\end{align}
	Application of this formula to the present Hamiltonian results in
	\begin{align}
		\chi
		&=-\frac{4e^2\gamma^{4}}{c^2} T\sum_{n, \bk}
		\left[
		\frac{1}{\left( \tilde{\ve}_n^2 + E_k^2 \right)^2}
		-\frac{8\gamma^4 k_x^2 k_y^2}{\left( \tilde{\ve}_n^2 + E_k^2 \right)^4}
		\right]
		\label{Chi}
		\\
		&=\frac{e^2 |\gamma|}{6c^2 \pi^2}
		\int_{-\infty}^{\infty}\!\! d\ve f(\ve)
		\left[
		\frac{1}{\sqrt{\ve_{+}^2 -\D^2}}
		-\frac{1}{\sqrt{\ve_{-}^2 -\D^2}}
		\right] .
		\label{Chi2}
\end{align}
	%
	%
	Despite their different starting points (eqs. (\ref{Phi}) and (\ref{Chi})), the final expressions are equivalent; an astonishing result.
	As a consequence, we obtain the following relation between the SHC and the orbital susceptibility in the insulating region :
	\begin{align}
		\sigma_{{\rm s}xy}
		=\frac{3m_{\rm e}c^{2}}{2e} \chi.
		\label{relation}
\end{align}
	In the insulating region, $\sigma_{xx}$ is suppressed, so that there are no dissipative current.
	Only in such a case, the SHE becomes dissipationless, and it becomes exactly the same as the dissipationless diamagnetic current.
	We can say that the spin-Hall current under an electric field appears as the diamagnetic current under a magnetic field.
	This correspondence would be due to the duality between the electricity and magnetism.
	%
	%
	%
	
	The relationship between the spin Hall conductivity and the spin density has been argued in the context of the two-dimensional quantum SHE of insulators for a non-relativistic model\cite{Yang2006}.
	According to them, the ``spin conserved'' part of the spin Hall conductivity in the insulating case is given by a St\v{r}eda-like formula: $\sigma_{{\rm s}yx}^{\rm II, (c)}=-\partial S_{z}/ \partial B $, where $S_{z}$ is the $z$-component spin density.
	Actually, in the present case, we find that the field derivative of the expectation value of magnetic moment $\mu_{{\rm e}z}$ (eq. (\ref{moment})) is proportional to $\sigma_{{\rm s}yx}$ as 
	\begin{align}
	\sigma_{{\rm s}yx} \propto -\partial \langle \mu_{{\rm s}z}\rangle /\partial B.
	\label{sigmom}
	\end{align}
	This is consistent with the claim of Ref. \citen{Yang2006}.
	However, $\langle \mu_{{\rm s}z}\rangle$ is not a measurable quantity and the relation in eq. (\ref{sigmom}) cannot be checked experimentally.
	Instead, eq. (\ref{relation}) can be examined experimentally.
	We also note that the St\v{r}eda-like formula used in Ref. \citen{Yang2006} is valid for the insulating case, while our calculation based on the Kubo formula is valid both for insulating and metallic case.
	%

\begin{figure}
\begin{center}
\includegraphics[width=7cm]{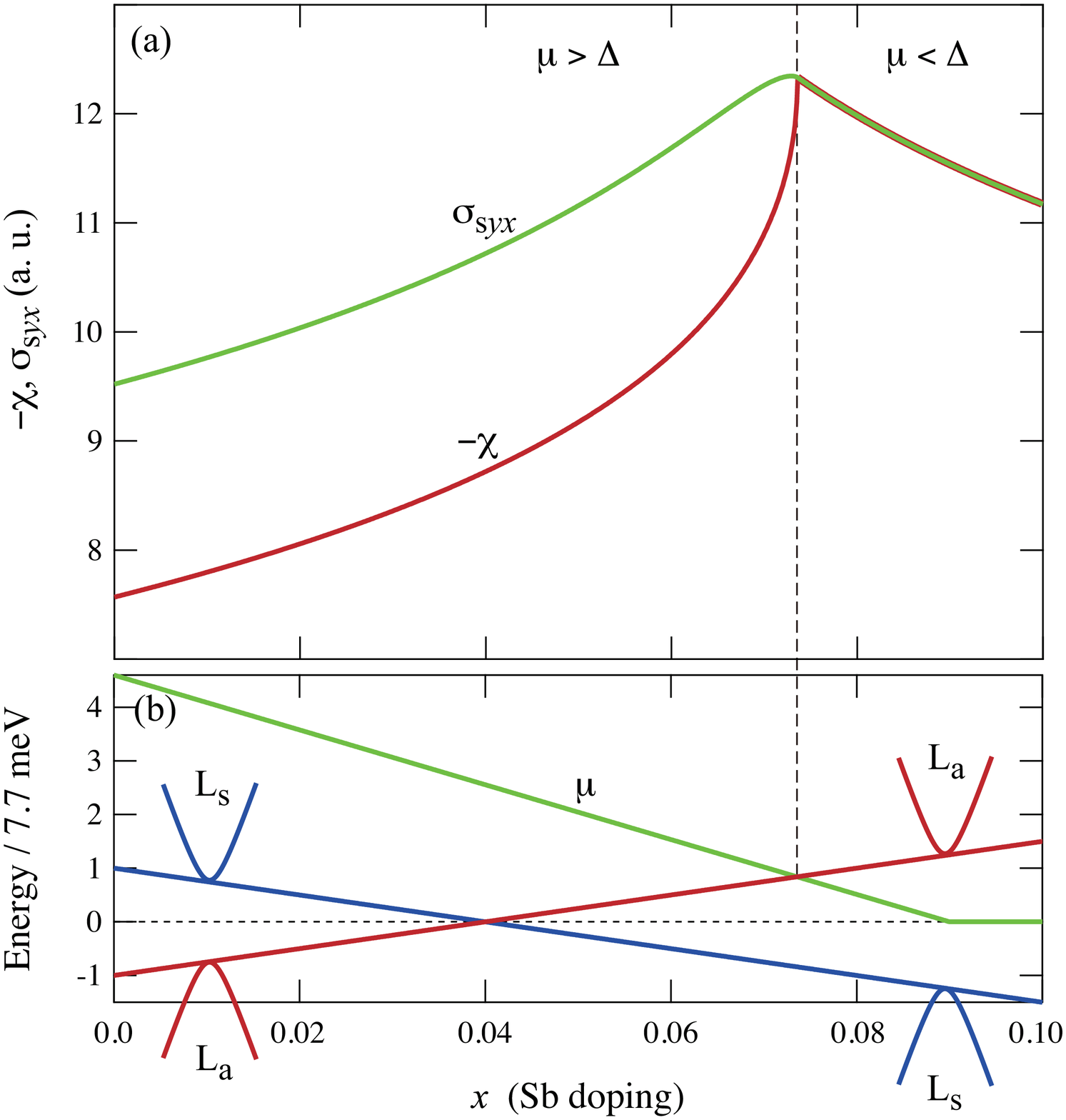}
\end{center}
\caption{(Color online) Antimony doping, $x$, dependence of (a) $\sigma_{{\rm s}yx}(x)$ and $-\chi(x)$, and (b) electron bands at $L$-point and $\mu$ of bismuth. The form of $\D(x)$ and $\mu (x)$ are defined in the text.}
\label{f3}
\end{figure}
	%
	Lastly, we discuss the implications of the present results to the experiments on bismuth.
	The band gap of electrons at $L$ points is $\D = 7.7$ meV, the chemical potential is $\mu =35.3$ meV, so that $\mu/\D = 4.6$ for pure semimetallic bismuth\cite{Smith1964,Zhu2011}.
	The band cut-off would be $E_{\rm c}=1$-2 eV\cite{Liu1995}, namely, $E_{\rm c}/\D = 130$-260.
	Substituting bismuth with antimony (Bi$_{1-x}$Sb$_x$) can change the band structure as depicted in Fig. \ref{f3} (b).\cite{Lenoir1996}
	By this substitution, the hole valence band at $T$ point is lowered, resulting in the decrease of $\mu$.
	The gap closes when $x\simeq 0.04$, where the bonding ($L_{\rm s}$) and antibonding ($L_{\rm a}$) bands are inverted, i.e., the topology of the bands changes from trivial to non-trivial\cite{Fu2007,Hsieh2008}.
	At $x\simeq 0.07$, the overlap between the electron conduction bands and the hole valence band vanishes, so that $\mu$ reaches the band-edge, i.e., $\mu(x)=\D(x)$.
	At $x\simeq 0.09$, the top of the hole valence band become lower than that of electron valence band, so that $\mu(x)=0$ beyond this composition.
	This doping dependence of $\mu$ and $\D$ can be simulated by $\pm \D(x)=1-x/0.04$ and $\mu (x)=4.6-4.6x/0.09$. 
	By substituting these $x$-dependent $\D (x)$ and $\mu(x)$ into eqs. (\ref{Csxy}) and (\ref{Chi2}), the doping dependence of $\chi (x)$ and $\sigma_{{\rm s}yx} (x)$ are obtained as is shown in Fig. \ref{f3} (a).

	The magnitude of $\chi(x)$ increases logarithmically, and a kink structure appears at $x=0.074$ where $\mu$ reaches the band-edge.
	Beyond this composition, the magnitude of $\chi(x)$ slightly decreases due to the increase of $\D (x)$.
	These properties of $\chi(x)$ agree well with the experimental results\cite{Verkin1967,Wehrli1968}, indicating the validity of our theory.
	The magnitude of $\sigma_{{\rm s}yx}(x)$ exhibits a similar behavior to $\chi(x)$;
	it increase logarithmically, and also have a kink at $x=0.074$.
	There is no anomaly at around $x=0.04$ at which the gap vanishes and the material is believed to change from a simple insulator to a topological insulator\cite{Fu2007,Hsieh2008}.
	The difference between $\sigma_{{\rm s}yx}(x)$ and $\chi(x)$ is due to the intra-band contribution $K_{{\rm s}yx}^{\rm I}$ in $\sigma_{{\rm s}yx}$.

	When we argue the transport property of bismuth, it is important to consider the contributions from holes at $T$ point, whose contributions are neglected in the present theory.
	The effective model for holes have been described by a parabolic band model with a large spin-mass term\cite{Smith1964,Zhu2011}, which would be originated from the SO interaction. 
	From this hole band, a finite contribution is expected. 
	However, this contribution will be small, since the gap at $T$ point is much larger and the band cutoff would be much smaller than those of $L$ points, as is the case of the hole contribution to the diamagnetism\cite{Fukuyama1970}.
	%

	%
	
	In this paper, we have discussed the spin-Hall effect of the isotropic Wolff Hamiltonian, which is fully relativistic, on the basis of the Kubo formula.
	%
	%
	It has been shown that the spin-Hall effect appears in this Dirac electron system. 
	Especially, the spin-Hall effect becomes maximum in the insulating region, where the electric current hardly flows.
	Therefore, we can obtain the dissipationless spin-Hall current there.
	This spin-Hall conductivity originates only from the inter-band effect, while the Hall conductivity originate both from intra- (dominant) and inter-band (small) effects.

	It has been found that there are two distinct contributions, i.e., $K_{{\rm s}yx}^{\rm I}$, ``transport contribution'', and $K_{{\rm s}yx}^{\rm II}$, ``thermodynamic contribution'', to the SHE.
	%
	%
	The latter has exactly the same chemical-potential dependence as the orbital susceptibility (diamagnetism).
	Eqs. (\ref{K2}), (\ref{Chi2}) and (\ref{relation}) are the first proof that shows the definite relation between the spin Hall conductivity and the diamagnetism with the fully relativistic model.
	This astonishing correspondence strongly suggests the spin-Hall effect has the same nature as the diamagnetism through the inter-band effect associated with the spin-orbit interaction and the duality between electricity and magnetism.
	This would play a crucial role for understanding the long-standing problem: how the transport coefficients relate to the orbital susceptibility\cite{Kubo1970}.

	The present work gives a fully relativistic theory of the spin-Hall effect not only in insulating but also in conducting state on equal footing for the first time.
	Also, our results have disclosed that the Dirac electron system will provide another ideal situation for investigating the spin-Hall effect, since it is very simple and gives not only qualitative but also quantitative agreements with experiments on bismuth.
	%

	\acknowledgement 
	We thank M. Shiraishi, E. Shikoh, K. Miyake and H. Kohno for helpful comments.
	This work is financially supported by Grant-in-Aid for Scientific Research (A) (No. 24244053) and for Young Scientists (B) (No. 23740269) from Japan Society for the Promotion of Science, and by MRL System of Osaka University.

\bibliographystyle{jpsj}
\bibliography{Bismuth}

\begin{thebibliography}{10}

\bibitem{Thomson1857}
W.~Thomson: Proc. R. Soc. {\bfseries 8} (1857) 546.

\bibitem{Binasch1989}
G.~Binasch, P.~Gr{\"u}nberg, F.~Saurenbach, and W.~Zinn: Phys. Rev. B
  {\bfseries 39} (1989) 4828.

\bibitem{Karplus1954}
R.~Karplus and J.~M. Luttinger: Phys. Rev. {\bfseries 95} (1954) 1954.

\bibitem{Nagaosa2010}
N.~Nagaosa, J.~Sinova, S.~Onoda, A.~H. MacDonald, and N.~P. Ong: Rev. Mod.
  Phys. {\bfseries 82} (2010) 1539.

\bibitem{Nagaosa2008}
N.~Nagaosa: J. Phys. Soc. Jpn. {\bfseries 77} (2008) 031010.

\bibitem{Dyakonov2008}
M.~I. Dyakonov and A.~V. Khaetskii: In M.~I. Dyakonov (ed), {\em Spin Physics
  in Semiconductors}, 2008.

\bibitem{Murakami2003}
S.~Murakami, N.~Nagaosa, and S.~C. Zhang: Science {\bfseries 301} (2003) 1348.

\bibitem{Sinova2004}
J.~Sinova, D.~Culcer, Q.~Niu, N.~A. Sinitsyn, T.~Jungwirth, and A.~H.
  MacDonald: Phys. Rev. Lett. {\bfseries 92} (2004) 126603.

\bibitem{Murakami2004}
S.~Murakami, N.~Nagaosa, and S.~C. Zhang: Phys. Rev. Lett. {\bfseries 93}
  (2004) 156804.

\bibitem{Kontani2007}
H.~Kontani, M.~Naito, D.~S. Hirashima, K.~Yamada, and J.~Inoue: J. Phys. Soc.
  Jpn. {\bfseries 76} (2007) 103702.

\bibitem{Kontani2008}
H.~Kontani, T.~Tanaka, D.~S. Hirashima, K.~Yamada, and J.~Inoue: Phys. Rev.
  Lett. {\bfseries 100} (2008) 096601.

\bibitem{Fukuyama1970}
H.~Fukuyama and R.~Kubo: J. Phys. Soc. Jpn. {\bfseries 28} (1970) 570.

\bibitem{Dresselhaus1971}
M.~S. Dresselhaus: J. Phys. Chem. Solids {\bfseries 32} (1971) 3.

\bibitem{Smith1964}
G.~E. Smith, G.~A. Baraff, and J.~M. Rowell: Phys. Rev. {\bfseries 135} (1964)
  A1118.

\bibitem{Zhu2011}
Z.~Zhu, B.~Fauqu\'e, Y.~Fuseya, and K.~Behnia: Phys. Rev. B {\bfseries 84}
  (2011) 115137.

\bibitem{Wolff1964}
P.~A. Wolff: J. Phys. Chem. Solids {\bfseries 25} (1964) 1057.

\bibitem{Kubo1970}
R.~Kubo and H.~Fukuyama: In E.~P. Keller, J.~C. Hensel, and F.~Stern (eds),
  {\em Proceedings of the Tenth International Conference on the Physics of
  Semiconductors}, 1970.

\bibitem{Fuseya2009}
Y.~Fuseya, M.~Ogata, and H.~Fukuyama: Phys. Rev. Lett. {\bfseries 102} (2009)
  066601.

\bibitem{Fuseya2012}
Y.~Fuseya, M.~Ogata, and H.~Fukuyama: J. Phys. Soc. Jpn., {\bfseries 81} (2012)
  013704.

\bibitem{Cohen1960}
M.~H. Cohen and E.~I. Blount: Phil. Mag. {\bfseries 5} (1960) 115.

\bibitem{Fukuyama1971}
H.~Fukuyama: Prog. Theor. Phys. {\bfseries 45} (1971) 704.

\bibitem{Yang2006}
M.-F. Yang and M.-C. Chang: Phys. Rev. B {\bfseries 73} (2006) 073304.

\bibitem{Liu1995}
Y.~Liu and R.~E. Allen: Phys. Rev. B {\bfseries 52} (1995) 1566.

\bibitem{Lenoir1996}
B.~Lenoir, M.~Cassart, J.-P. Michenaud, H.~Scherrer, and S.~Scherrer: J. Phys.
  Chem. Solids {\bfseries 57} (1996) 89.

\bibitem{Fu2007}
L.~Fu and C.~L. Kane: Phys. Rev. B {\bfseries 76} (2007) 045302.

\bibitem{Hsieh2008}
D.~Hsieh, D.~Qian, L.~Wray, Y.~Xia, Y.~S. Hor, R.~J. Cava, and M.~Z. Hasan:
  Nature {\bfseries 452} (2008) 970.

\bibitem{Verkin1967}
B.~Verkin, L.~B. Kuz'micheva, and I.~V. Svechkarev: JETP Letters {\bfseries 6}
  (1967) 225.

\bibitem{Wehrli1968}
L.~Wehrli: Phys. Kondens. Materie {\bfseries 8} (1968) 87.

\end{thebibliography}

\end{document}